\documentclass[aps,prb,bibliography,superscriptaddress,twocolumn]{revtex4-2}

\usepackage[utf8]{inputenc}
\usepackage{amsmath,amssymb,amsfonts,mathtools}
\usepackage{physics}
\usepackage{braket}
\usepackage{bm}
\usepackage{graphicx}
\usepackage{subfigure}
\usepackage{textcomp}
\usepackage{color}
\usepackage{float}
\usepackage[dvipsnames]{xcolor}
\usepackage{mathrsfs}
\usepackage{natbib}
\usepackage{makecell}
\usepackage{soul}
\usepackage{multirow}

\newcommand{\bk}{\mathbf{k}}

\begin{document}

\title{Bicircular light tunable topological phases and nonlinear Hall responses in odd-parity magnets}

\author{Md Afsar Reja}
\email{afsarmd@iisc.ac.in}
\affiliation{Solid State and Structural Chemistry Unit, Indian Institute of Science, Bangalore 560012, India}

\author{Awadhesh Narayan}
\email{awadhesh@iisc.ac.in}
\affiliation{Solid State and Structural Chemistry Unit, Indian Institute of Science, Bangalore 560012, India}

\date{\today}
\begin{abstract} Altermagnets are an emerging class of magnetic materials that exhibit characteristic features of both ferromagnets and antiferromagnets. Odd-wave magnets, where the momentum-dependent spin splitting follows an odd-parity form factor, provide a distinct platform for exploring unconventional magnetic and topological phenomena. In this work, we demonstrate light-induced topological phase transitions in odd-parity magnets driven by bicircular light. We show that a cascade of topological phase transitions occur as the relative amplitude ratio of the two beams is varied. These are accompanied by sharp changes of the quantum metric. We demonstrate that bicircular light unlocks nonlinear Hall responses arising from Berry curvature and quantum metric in irradiated odd-wave magnets, and formulate the underlying selection rules. Our results show that both the topology and the nonlinear responses in odd-parity magnets can be directly controlled by a bicircular drive.
\end{abstract}
\maketitle

\section{Introduction}

Altermagnets are a class of magnetic materials that feature spin-split electronic bands, similar to ferromagnets, while retaining zero net magnetization. In altermagnets, opposite-spin sublattices are related by crystal rotational or mirror symmetries rather than by translational or inversion symmetries, which makes them distinct from both ferromagnets and conventional antiferromagnets~\cite{vsmejkal2022beyond,vsmejkal2022emerging,hayami2019momentum,yuan2020giant}. Over the past few years,  altermagnets have attracted considerable attention as they provide a versatile platform for exploring a wide range of physical phenomena, including topological phases, multiferroicity, and unconventional superconductivity, while also offering promising opportunities for potential spintronic applications~\cite{song2025altermagnets,tamang2025altermagnetism,devaraj2026altermagnetism,jungwirth2026altermagnetic}.

Depending on the parity of the momentum-dependent spin splitting, unconventional magnets are classified into even-wave (\(d\)-, \(g\)-wave) and odd-wave (\(p\)-, \(f\)-wave) classes. Even-wave altermagnets typically break time-reversal symmetry, whereas odd-waves preserve effective time reversal but break inversion symmetry, leading to distinct momentum-dependent band splittings~\cite{hellenes2023p,brekke2024minimal,mitscherling2026microscopic,sim2026quantum}. The different types of inversion symmetry breaking in $p$-wave magnets leads to different kinds of spin-orbit coupling effects~\cite{fakhredine2026interplay,liu2026nonrelativistic}. A nonrelativistic Edelstein effect has been predicted from first-principles calculations in a $p$-wave candidate material~\cite{chakraborty2025highly}. A variety of non-Hermitian phenomena, including the emergence of symmetry-protected exceptional rings, have also been explored in odd-wave magnets~\cite{reja2025neel,alipourzadeh2026opto,correa2026emergent}. More recently, electrical switching of $p$-wave magnets has been demonstrated experimentally~\cite{song2025electrical,yamada2025metallic}.
 
Floquet engineering is, by now, an established framework to control quantum phases~\cite{oka2019floquet,rudner2020band,sen2021analytic}. It has been proposed to tailor the properties of topological matter~\cite{oka2009photovoltaic,lindner2011floquet,cayssol2013floquet,narayan2015floquet,yan2016tunable,narayan2016tunable, wang2023floquet}, non-Hermitian systems~\cite{wu2020floquet,banerjee2020controlling,zhou2023non,dash2024floquet}, and quantum magnets~\cite{owerre2017floquet,chaudhary2019orbital,li2024floquet}, to highlight just a few. In recent years, the exploration of Floquet engineering using bicircularly polarized light, utilizing two circularly polarized light beams, has been rapidly evolving. With multiple control knobs, it provides a greater flexibility in tuning the symmetries and the resulting response of a diverse array of systems. Its proposed applications range from topological phase transitions~\cite{trevisan2022bicircular,wang2023topological,ganguli2025floquet,chowdhury2025phase}, photocurrents~\cite{ikeda2023photocurrent,ikeda2024controllable,kanega2025two}, as well as quantum geometry~\cite{guo2025bicircular,yuan2026quantum,zhu2026floquet}, among several others.   

\begin{figure}[b]
\centering
\includegraphics[width=\columnwidth]{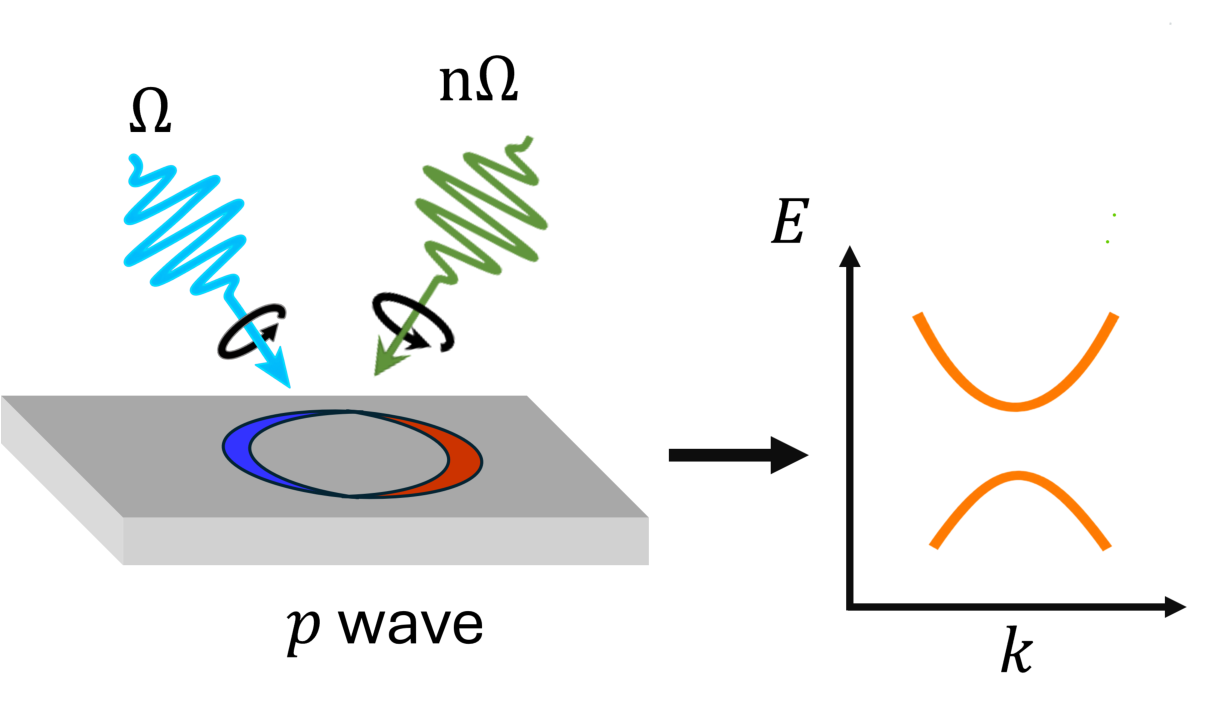}
\caption{\textbf{Schematic of the illuminated odd-parity magnet setup.} Bicircular light drives topological phase transitions in odd-wave magnets and simultaneously enables tunable nonlinear Hall responses. The counter-rotating $\Omega-n\Omega$ bicircular light is shown illuminating a $p$-wave magnet as a representative.}
\label{fig:schematic}
\end{figure}

In the context on unconventional magnets, using light has very recently emerged as a promising approach for controlling both even- and odd-wave magnetic phases. For example, monochromatic irradiation has been shown to tune topological phases and generate persistent spin textures~\cite{ghorashi2025dynamical}, while periodic driving has been shown to induce odd-frequency spin-triplet superconducting correlations in unconventional magnets~\cite{fu2026floquet}. Additionally, Floquet engineering of even-wave altermagnets has been proposed in diverse settings~\cite{ganguli2026tunable,yarmohammadi2026giant,iurov2026exploring,cheraghchi2026floquet,li2026tunable,liu2026linearly,yarmohammadi2026spin,shah2026floquet,qin2026anomalous,yarmohammadi2026efficient,fu2026complete}. Furthermore, Floquet driving has been proposed as a route to generate odd-wave magnetic states from conventional antiferromagnetic systems, highlighting the potential role of light in engineering odd-wave magnets~\cite{huang2026light,li2026floquet,zhu2026floquet,liu2026light}.

Despite these developments, the interplay of bicircular driving, topology, quantum geometry, and nonlinear Hall transport in odd-parity $p$- and $f$-wave magnets remains unexplored. In this work, we investigate the Floquet engineering of odd-parity $p$- and $f$-wave magnets using counter-rotating bicircular light, as
illustrated schematically in Fig.~\ref{fig:schematic}. For a square-lattice $p$-wave model, we show that the relative amplitude ratio of the two driving components provides direct control over the light-induced mass and the resulting Chern phases. We find that the associated topological transitions are further reflected in a pronounced enhancement of the quantum weight near the phase boundaries. We then show that bicircular driving generates tunable nonlinear Hall
responses governed by the Berry curvature dipole and Berry connection polarizability, and identify an odd-even selection rule controlled by the frequency ratio of the drive. Extending the analysis to a triangular-lattice $f$-wave magnet, we obtain a richer Chern phase diagram and find a distinct
frequency-ratio selection rule associated with the underlying threefold lattice symmetry. We additionally demonstrate that the momentum-resolved spin texture can be manipulated through the properties of the bicircular field.

The remainder of the paper is organized as follows. Section~\ref{sec: Topological_phase_p} discusses the Floquet topological phases and quantum geometry of the $p$-wave model. In Section~\ref{sec:response_p}, we present its nonlinear responses and spin textures. In Section~\ref{sec:f_wave} we extend the analysis to the $f$-wave model. Finally, Section~\ref{sec:summary} summarizes our results and outlook.

\section{Bicircular light driven topological phases in $p$-wave magnets}

We begin by considering a square-lattice $p$-wave magnet described by the Hamiltonian~\cite{brekke2024minimal,hellenes2023p,ezawa2025third},

\begin{equation}
\begin{aligned}
H_p(\bk)
&=
\left[-2t(\cos k_x+\cos k_y)-\mu\right]\sigma_0
+\lambda\left(\sin k_y\,\sigma_x-\sin k_x\,\sigma_y\right) \\
&\quad
+J_p\left(c_p\sin k_x+s_p\sin k_y\right)\sigma_z .
\end{aligned}
\label{eq:static_pwave}
\end{equation}

Here $t$ denotes the nearest-neighbor hopping amplitude, $\mu$ is the chemical potential, $\lambda$ represents the Rashba spin-orbit coupling strength, $J_p$ is the $p$-wave exchange coupling, and $c_p=\cos\theta_p$ and $s_p=\sin\theta_p$, with $\theta_p$ specifying the orientation of the $p$-wave magnetic order. We set the lattice constant to unity. Writing $H_p(\bk)=d^0(\bk)\sigma_0+\mathbf{d}(\bk)\cdot\bm{\sigma}$, the $\mathbf{d}(\bk)$ vector is given by 

\begin{equation}
\mathbf{d}(\bk)=\left(\lambda\sin k_y,-\lambda\sin k_x,J_p(c_p\sin k_x+s_p\sin k_y)\right).
\end{equation}

We introduce the counter-rotating $\Omega$-$n\Omega$ bicircular light through the Peierls substitution $\bk\rightarrow\bk+\mathbf{a}(\tau)$, where $\tau=\Omega t$. The components of the vector potential are chosen as

\begin{equation}
\begin{aligned}
a_x(\tau) &= A_1\cos\tau+A_n\cos(n\tau-\alpha), \\
a_y(\tau) &= A_1\sin\tau-A_n\sin(n\tau-\alpha),
\end{aligned}
\label{eq:bcl_drive}
\end{equation}

where $A_1$ and $A_n$ denote the amplitudes of the fundamental and higher-harmonic component, respectively, and $\alpha$ is their relative phase. We parameterize the two amplitudes in terms of the total field strength, $R$, and the relative amplitude, $r$, as

\begin{equation}
A_1=\frac{R}{\sqrt{1+r^2}},\qquad
A_n=\frac{rR}{\sqrt{1+r^2}},\qquad
r=\frac{A_n}{A_1},
\label{eq:Rr_param}
\end{equation}

such that $A_1^2+A_n^2=R^2$ for fixed $R$.

Under the drive, the time-dependent Hamiltonian becomes

\begin{equation}
\begin{aligned}
H_p(\bk,\tau)
&=
\left[-2t\left\{\cos(k_x+a_x)+\cos(k_y+a_y)\right\}-\mu\right]\sigma_0 \\
&\quad
+\lambda\left[\sin(k_y+a_y)\sigma_x-\sin(k_x+a_x)\sigma_y\right] \\
&\quad
+J_p\left[c_p\sin(k_x+a_x)+s_p\sin(k_y+a_y)\right]\sigma_z .
\end{aligned}
\label{eq:driven_pwave}
\end{equation}

To first order in the high-frequency expansion, the effective Hamiltonian reads~\cite{oka2019floquet,rudner2020band,sen2021analytic}

\begin{equation}
H_{\mathrm{eff}}(\bk)=H_0(\bk)+\sum_{m=1}^{\infty}\frac{[H_m(\bk),H_{-m}(\bk)]}{m\hbar\Omega}+\mathcal{O}(\Omega^{-2}),
\label{eq:vanVleck}
\end{equation}

where the Fourier components are of the form,

\begin{equation}
H_m(\bk)=\frac{1}{2\pi}\int_0^{2\pi}H_p(\bk,\tau)e^{-im\tau}\,d\tau.
\label{eq:fourier_blocks}
\end{equation}

\begin{figure}[t]
\centering
\includegraphics[width=\columnwidth]{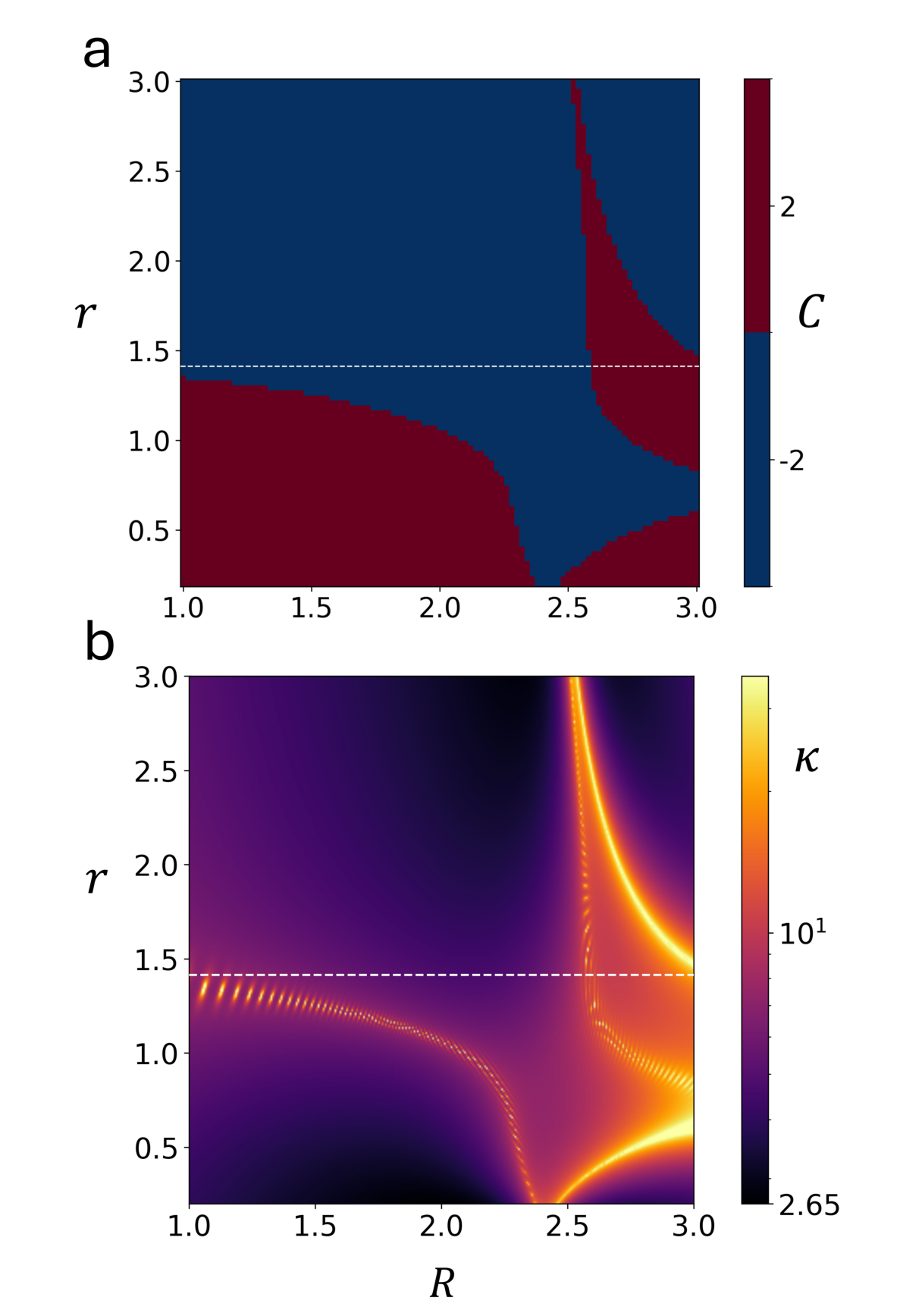}
\caption{\textbf{Chern number and quantum weight for the $p$-wave magnet under bicircular drive.} (a) Topological phase diagram showing Chern number as a function of the total light amplitude $R$ and the relative amplitude ratio $r$. (b) The quantum weight is presented in the same phase space. We find that the quantum weight satisfies the geometric bound throughout the $(R,r)$ parameter space, reaching large values near the phase boundaries. The other parameters are chosen to be $n=2$, $\alpha=0$, $t=0.1$, $\mu=-1.0$, $\lambda=0.1$, $J_p=0.3$, $\theta_p=\pi/4$, and $\Omega=3.0$.}
\label{fig:pwave_chern}
\end{figure}

We begin with the low-energy Hamiltonian to obtain analytical insights into the Floquet mass and associated gap closing, and then turn to the full lattice effective Hamiltonian to determine the complete topological phase diagram. Under bicircular driving, the $\mathbf{d}(\bk)$ vector of the low-energy Hamiltonian becomes,
$\left(\lambda(k_y+a_y),\,-\lambda(k_x+a_x),\,J_p[c_p(k_x+a_x)+s_p(k_y+a_y)]\right)$.
To first order in the Floquet-Magnus expansion, the effective Hamiltonian components can be written compactly as,

\begin{equation}
\begin{pmatrix}
\lambda k_y+\dfrac{\lambda J_p s_p\Phi_n}{\hbar\Omega} \\[6pt]
-\lambda k_x-\dfrac{\lambda J_p c_p\Phi_n}{\hbar\Omega} \\[6pt]
J_p(c_pk_x+s_pk_y)-\dfrac{\lambda^2\Phi_n}{\hbar\Omega}
\end{pmatrix}.
\label{eq:p_deff}
\end{equation}

Here, we define,

\begin{equation}
\Phi_n
=
A_1^2-\frac{A_n^2}{n}
=
\frac{R^2}{1+r^2}
\left(1-\frac{r^2}{n}\right).
\label{eq:chi_n}
\end{equation}

For details of this calculation please see the Appendix~\ref{sec:floquet_formalism}. The effective Floquet mass is proportional to $\Phi_n$, and therefore changes sign when $\Phi_n=0$, corresponding to $r=\sqrt{n}$. For an $\Omega$-$2\Omega$ drive, this condition reduces to $r=\sqrt{2}$. The low-energy theory therefore captures the Floquet mass reversal and the associated gap-closing condition, while the corresponding Chern number, $C$, is determined from the full lattice model. 

\label{sec: Topological_phase_p}
\begin{figure*}[t]
\centering
\includegraphics[width=0.7\textwidth]{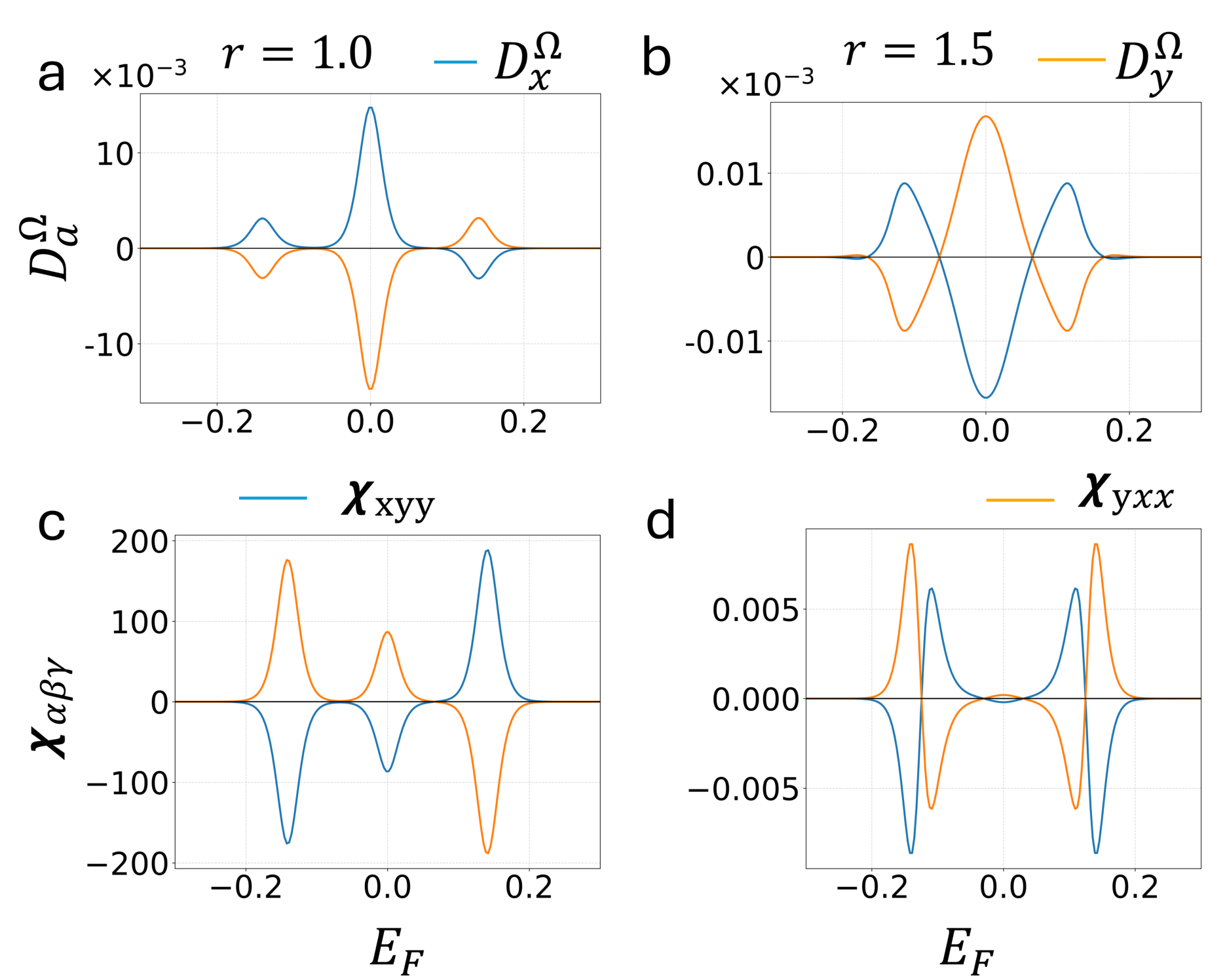}
\caption{\textbf{Nonlinear responses of the bicircular light driven $p$-wave magnet.} Berry curvature dipole response for (a) $r=1.0$ and (b) $r=1.5$. The blue and orange curves show $D_{x}^{\Omega}$ and $D_{y}^{\Omega}$, respectively. Berry connection polarizability response for (c) $r=1.0$ and (d) $r=1.5$. Here the blue and orange curves indicate $\chi_{xyy}$ and $\chi_{yxx}$ components, respectively. We note that the relative amplitude ratio of the bicircular drive, $r$, acts as a tuning knob for the nonlinear responses. We keep the remaining parameters fixed at  $t=0.1$, $\lambda=0.1$, $J_p=0.3$, $\theta_p=\pi/4$, $\alpha=\pi/4$, $n=2$, $R=2$, and $\Omega=3.0$.}
\label{fig:pwave_BCD_response}
\end{figure*}

We construct the light-dressed effective Hamiltonian by retaining Floquet harmonics up to $m=10$. We have verified numerically that including additional harmonics does not appreciably modify the effective Hamiltonian within the parameter regime that we consider here. In Fig.~\ref{fig:pwave_chern}(a), we present the complete topological phase diagram in the $(R,r)$ plane for the $\Omega$-$2\Omega$ drive for the full lattice Hamiltonian. A clear topological transition appears near $r=\sqrt{2}$ condition indicated by the white dashed line, in agreement with our low-energy prediction. Away from this regime, lattice effects and contributions from higher Floquet harmonics modify the effective Hamiltonian and lead to a richer phase diagram. Under bicircular driving, the first-order Floquet correction opens gaps at the high symmetry points and generates a light-induced mass controlled by $\Phi_n$. The mass has the same sign relative to the local chirality at all four high symmetry points, so their half-integer Chern contributions add constructively. As a result, the driven $p$-wave model realizes topological phases with $C=\pm2$. As we noted, the sign reversal occurs when $\Phi_n$ changes sign.

At relatively low driving amplitudes $R$, the topological phases are governed primarily by the amplitude ratio $r$, with the phase boundary remaining close to an approximately constant value of $r$. As $R$ is increased, the boundary gradually bends toward smaller $r$, indicating a stronger dependence on the total driving amplitude. Around $R\simeq 2.5$, the phase diagram develops a pronounced re-entrant structure, where the phase boundary turns back toward larger $r$ and additional topological regions emerge. This evolution highlights the combined role of $R$ and $r$ in controlling the light-induced Chern phases. We note that monochromatic circularly polarized light, obtained by setting $A_n=0$, i.e. $r=0$, can also induce a topological phase through the generation of a nonzero Floquet mass. However, bicircular driving leads to a richer topological phase diagram with notable re-entrant structure and, as discussed in the following sections, also provides a route to generate and control nonlinear Hall responses.

\begin{figure*}[t]
\centering
\includegraphics[width=0.75\textwidth]{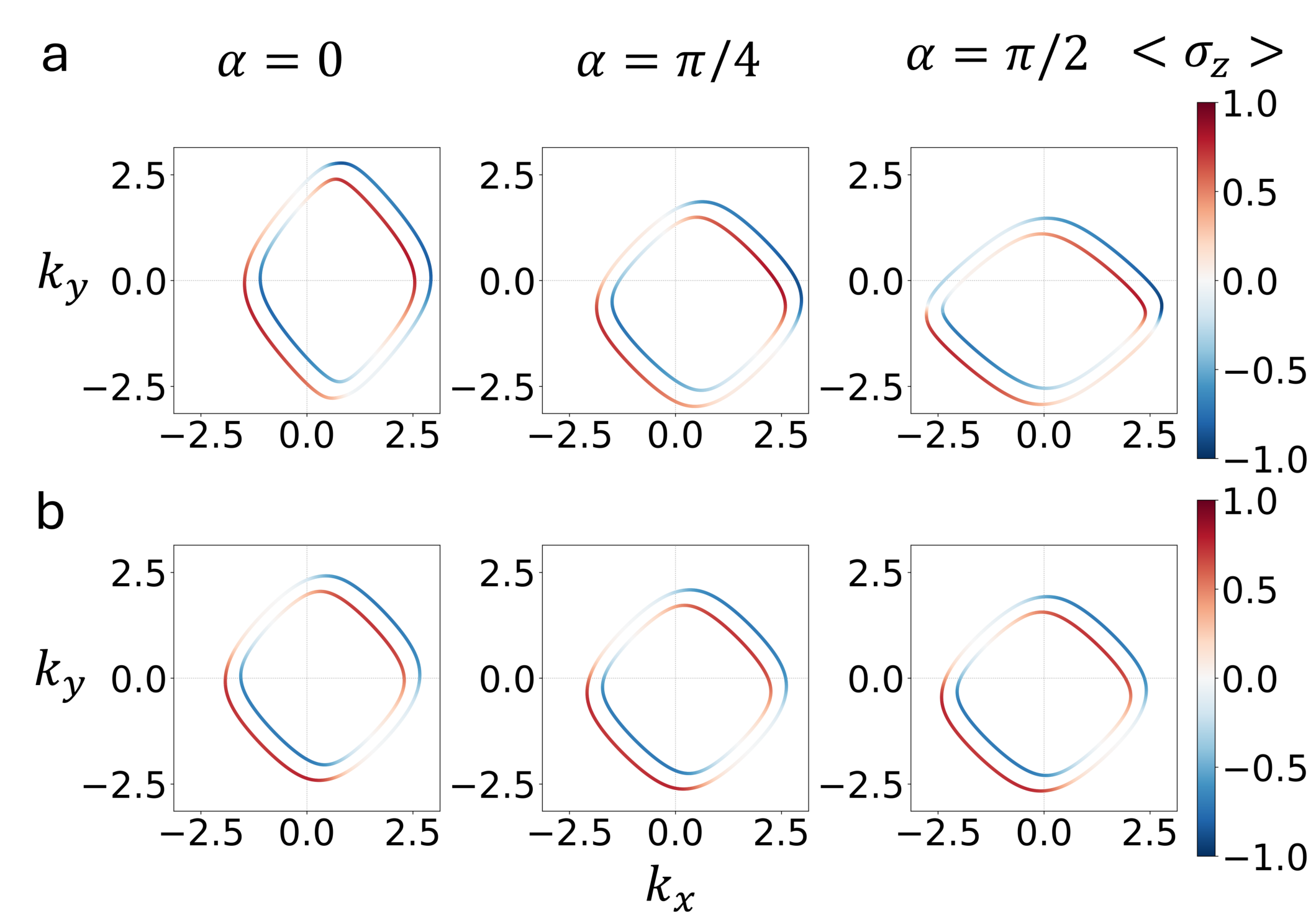}
\caption{\textbf{Spin texture of the driven $p$-wave magnet.} The expectation value of the $z$ component of the spin is plotted across $k_x-k_y$ for (a) $r=0.5$ and (b) $r=2.0$. The three columns show $\alpha=0$, $\pi/4$, and $\pi/2$, respectively. We find that the spin texture can be tuned by the bicircular light parameters, in particular the relative amplitude ratio $r$ and the relative phase $\alpha$. Other parameters were chosen to be $t=1$, $\mu=-0.250$, $\lambda=0.3$, $J_p=0.3$, $\theta_p=\pi/4$, $\Omega=3.0$, $n=3$, and $R=2$.}
\label{fig:pwave_spin_texture}
\end{figure*}

The quantum metric, $g$, defined as the real part of the quantum geometric tensor, plays an important role in characterizing the geometry of Bloch states~\cite{provost1980riemannian,torma2023essay,liu2025quantum,yu2025quantum,gao2508quantum}. Furthermore, it governs a variety of material properties and quantum-geometric responses. Here, we characterize the topological phases obtained by bicircular drive by evaluating the Brillouin-zone-integrated quantum metric or the quantum weight~\cite{onishi2025quantum}, 

\begin{equation}
\mathcal{K}=\frac{1}{2\pi}\int_{\mathrm{BZ}}\mathrm{Tr}\,g\,d^2k.
\end{equation}

The quantum weight $\mathcal{K}$ in the $(R,r)$ plane is shown in Fig.~\ref{fig:pwave_chern}(b). Across the phase diagram, we find that the quantum weight satisfies the geometric bound $\mathcal{K}\geq |C|$, consistent with the relation between the quantum metric and the Chern number for an isolated band~\cite{ozawa2021relations}. The quantum weight map in Fig.~\ref{fig:pwave_chern}(b) closely follows the Chern-number phase boundaries, exhibiting pronounced enhancement in their vicinity, consistent with the topological phase transitions. Next, we show how the application of bicircular light can unlock nonlinear responses in $p$-wave magnets.

\section{Tunable nonlinear responses in $p$-wave magnets} 
\label{sec:response_p}

To understand the nonlinear Hall response of the $p$-wave magnets under illumination, we first construct the effective light-dressed Hamiltonian as prescribed in the previous section. From the resulting Floquet quasienergies and eigenstates, we evaluate the Berry curvature dipole (BCD) as~\cite{sodemann2015quantum,bandyopadhyay2024non},

\begin{equation}
D_a^{\Omega}
=
\sum_s\int_{\mathrm{BZ}}
\frac{d^2k}{(2\pi)^2}
\left(-\frac{\partial f}{\partial E_s}\right)
v_{s,a}(\bk)\Omega_s(\bk),
\label{eq:BCD}
\end{equation}

where $E_s(\bk)$ denotes the Floquet quasienergy, $v_{s,a}=\partial_{k_a}E_s$ is the band velocity, $f$ is the Fermi function, and $\Omega_s(\bk)$ is the Berry curvature of band $s$. We set $e=\hbar=1$ for simplicity. The intrinsic second-order Hall conductivity associated with the Berry connection polarizability (BCP) for $a\neq b$ is given by~\cite{liu2021intrinsic,wang2021intrinsic}

\begin{equation}
\chi^{\mathrm{int}}_{abb}
=
\sum_s
\int_{\mathrm{BZ}}
\frac{d^2k}{(2\pi)^2}
\left(-\frac{\partial f}{\partial E_s}\right)
\left[
v_{s,b}G^s_{ab}
-
v_{s,a}G^s_{bb}
\right].
\label{eq:bcp_abb}
\end{equation}

Here,

\begin{equation}
G^{s}_{ab}(\bk)
=
2\,\mathrm{Re}
\sum_{s'\neq s}
\frac{
\mathcal{A}^{ss'}_{a}(\bk)
\mathcal{A}^{s's}_{b}(\bk)
}{
E_s(\bk)-E_{s'}(\bk)
},
\label{eq:bcp_tensor}
\end{equation}

with $\mathcal{A}^{ss'}_{a}= \langle u_s|i\partial_{k_a}|u_{s'}\rangle$. For our square-lattice $p$-wave model, monochromatic circularly polarized light does not generate a finite BCD or BCP response. In evaluating the nonlinear responses, we assume a quasi-steady Floquet regime in which relaxation processes establish Fermi-Dirac-like occupations of the effective Floquet bands. The different components of the BCD response are presented in Fig.~\ref{fig:pwave_BCD_response}(a) for $r=1$ for applied bicircular light. In this case, we find that $D_x^\Omega$ and $D_y^\Omega$ exhibit approximately equal magnitudes with opposite signs. This symmetric behavior is generally lost when $\theta_p\neq\pi/4$ and when the relative phase $\alpha\neq\pi/4$. Upon increasing the amplitude ratio to $r=1.5$, the BCD response shape changes substantially and the overall magnitude is reduced, as shown in Fig.~\ref{fig:pwave_BCD_response}(b). The overall magnitude of the response can also be controlled by varying the driving amplitude $R$. We also observe a similar light tunability for the BCP responses, $\chi_{xyy}$ and $\chi_{yxx}$, as shown in Figs.~\ref{fig:pwave_BCD_response}(c) and \ref{fig:pwave_BCD_response}(d) for $r=1$ and $r=1.5$, respectively.

A characteristic feature of bicircular driving is that the nonlinear response depends strongly on the frequency ratio $n$ of the $\Omega$-$n\Omega$ field. An intriguing cancellation of the BCD and BCP response occurs for odd values of $n$, whereas finite responses become symmetry allowed for even $n$ only. This behavior originates from the half-period property of the bicircular field. For odd $n$, the vector potential satisfies $\mathbf{a}(\tau+\pi)=-\mathbf{a}(\tau)$, which imposes parity constraints on the Fourier coefficients entering $H_m$. As a consequence, the Floquet effective Hamiltonian retains a reciprocal quasienergy spectrum, $E_s(-\bk)=E_s(\bk)$, while the Berry curvature and quantum metric remain even under $\bk\rightarrow-\bk$. Since the band velocity is odd, $v_{s,a}(-\bk)=-v_{s,a}(\bk)$, the contributions from $\bk$ and $-\bk$ cancel in Eq.~\eqref{eq:BCD} and Eq.~\eqref{eq:bcp_abb}, yielding $D_a^{\Omega}=0$ and $\chi ^{int}_{abb}=0$ for both the monochromatic as well as the odd-$n$ cases. In contrast, for even $n$, the relation $\mathbf{a}(\tau+\pi)=-\mathbf{a}(\tau)$ is no longer satisfied. Consequently, the Fourier components do not have definite momentum parity, and the effective Hamiltonian can contain mixed even- and odd-in-momentum contributions. The cancellation between opposite momenta is therefore lifted, allowing finite Berry curvature- and quantum metric-induced nonlinear responses. Further details of this odd-even selection rule are provided in Appendix~\ref{sec:selection_rules}.

We can effectively tune both the BCD and BCP responses by varying the properties of the bicircular drive, such as the amplitude ratio $r$, the total amplitude $R$, and the relative phase $\alpha$. We find that both the magnitude and sign of the nonlinear response can be tuned by varying $\alpha$, reflecting the phase-dependent momentum-space asymmetry generated by the two-frequency drive. The response also depends on the parameter $\theta_p$ of the $p$-wave magnetic order, which controls the relative orientation between the magnetic spin texture and the light-induced Floquet terms. Thus, the frequency ratio $n$, relative phase $\alpha$, and magnetic orientation $\theta_p$ provide independent control parameters for tuning the nonlinear Hall response in irradiated $p$-wave magnets.

\begin{figure}[t]
\centering
\includegraphics[width=\columnwidth]{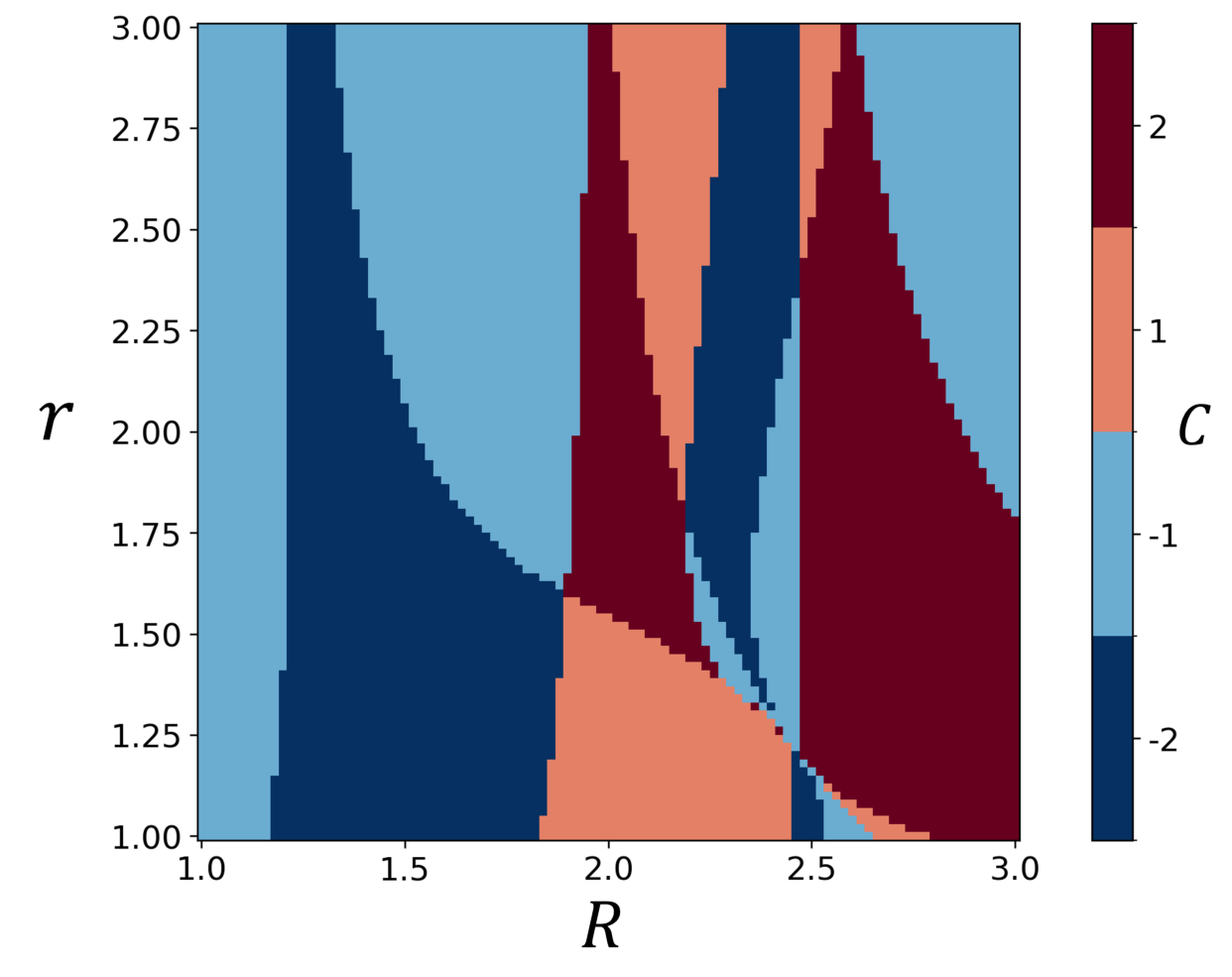}
\caption{\textbf{Chern number phase diagram of bicircular light illuminated $f$-wave magnet.} Bicircular light induces multiple topological phases characterized by $C=\pm1$ and $C=\pm2$. The phase boundaries are controlled by the total driving amplitude $R$ and the relative amplitude ratio $r$. Here we set other parameters as $n=2$, $\alpha=0$, $t=0.1$, $\mu=-1.0$, $\lambda=0.1$, $J_f=0.3$, and $\Omega=3.0$.}
\label{fig:fwave_chern}
\end{figure}

In Fig.~\ref{fig:pwave_spin_texture}, we show the momentum-resolved out-of-plane spin polarization $\langle\sigma_z\rangle$ on the Fermi contours for $\alpha_p=0$, $\pi/4$, and $\pi/2$, with the top and bottom panels corresponding to $r=0.5$ and $r=2.0$, respectively. We choose \(n=3\) here as a representative odd frequency ratio case to emphasize that the spin texture remains tunable even when the nonlinear Hall response is symmetry forbidden. Varying the amplitude ratio $r$ changes the relative weight of the two driving harmonics and consequently modifies both the Fermi contour geometry as well as the magnitude and sign distribution of $\langle\sigma_z\rangle$. Changing the relative phase $\alpha$ further alters the orientation and anisotropy of the spin texture through the interference between the two driving components. Thus, bicircular light provides a means of controlling the momentum-dependent spin polarization without modifying the underlying material parameters. Next, we generalize our results to higher harmonic odd-wave magnets, and present the effects of bicircular light on $f$-wave magnets.

\begin{figure*}[t]
\centering
\includegraphics[width=0.7\textwidth]{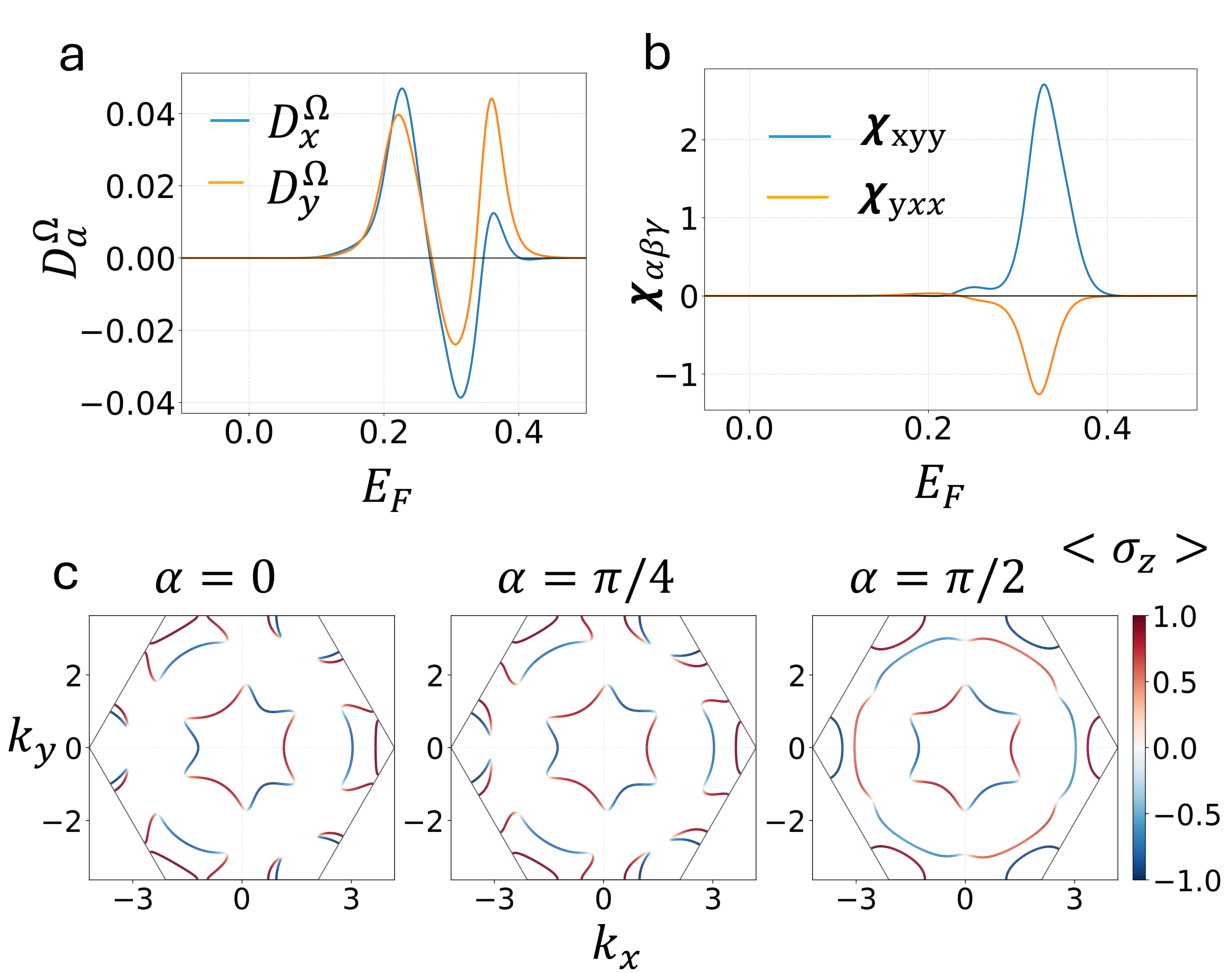}
\caption{\textbf{Nonlinear responses and spin texture of the light-driven $f$-wave magnet.} (a) Berry curvature dipole and (b) Berry connection polarizability components plotted as a function of the Fermi energy $E_F$. Both responses can be tuned by the relative phase $\alpha$, amplitude ratio $r$, and frequency ratio $n$ of the bicircular drive. Here we choose other parameters as $n=3$, $\alpha=\pi/4$, $t=0.1$, $\mu=0.0$, $\lambda=0.1$, $J_f=0.3$, $r=1.0$, $R=2.0$, and $\Omega=3.0$. (c) The spin texture is controlled by the relative phase $\alpha$ of the bicircular drive. The three panels correspond to $\alpha=0$, $\pi/4$, and $\pi/2$, demonstrating the pronounced phase dependence. Here we choose $n=2$, $t=0.1$, $\mu=0.30$, $\lambda=0.1$, $J_f=0.3$, $r=2.0$, $R=2.0$, and $\Omega=3.0$.}
\label{fig:response}
\end{figure*}

\section{Bicircular light illuminated $f$-wave magnets: topological phases and nonlinear responses} 
\label{sec:f_wave}

To analyze the effects of light illumination, we consider a triangular lattice $f-$wave magnet described by the following Hamiltonian~\cite{ezawa2025third},

\begin{equation}
H_f(\bk)
=
\epsilon_0(\bk)\sigma_0 +
\mathbf{d}_f(\bk)\cdot\bm{\sigma},
\label{eq:fwave_hamiltonian}
\end{equation}

where,

\begin{equation}
\mathbf d_f(\bk)
=
\left(
\begin{aligned}
&-\frac{\lambda}{\sqrt{3}}\left(\sin q_2+\sin q_3\right),\\
&\frac{\lambda}{3}\left(2\sin q_1+\sin q_2-\sin q_3\right),\\
&J_f\left(\sin 2q_1-\sin 2q_2+\sin 2q_3\right)
\end{aligned}
\right),
\end{equation}

with,

\begin{equation}
    \epsilon_0(\bk)=t\left[3-\cos q_1-\cos q_2-\cos q_3\right]-\mu,
\end{equation}

\begin{equation}
    q_1=k_x, \quad q_2=\frac{k_x+\sqrt{3}k_y}{2}, \quad q_3=\frac{-k_x+\sqrt{3}k_y}{2}.
\end{equation}

Here, $t$ denotes the nearest-neighbor hopping amplitude, $\lambda$ is the Rashba spin-orbit coupling strength, and $J_f$ represents the $f$-wave magnetic exchange coupling.

After performing a Floquet analysis analogous to that of the $p$-wave model, we find that the $f$-wave system exhibits a considerably richer Chern number phase diagram, as we present in Fig.~\ref{fig:fwave_chern}. In particular, the bicircularly driven $f$-wave magnet hosts several topological phases characterized by $C=\pm1$ and $C=\pm2$, whereas the $p$-wave model supports only the $C=\pm2$ phases. A cascade of topological phase transitions occur as either the relative light amplitude $r$ or the total light amplitude $R$ are tuned. 

A simple understanding of the richer Chern phase diagram follows from the low-energy theory at the symmetry-related band-touching points. In the triangular $f$-wave model, in addition to the Dirac point at $\Gamma$, there are three symmetry-related $M$ points. The linearized Hamiltonian has opposite chirality at these points, $\chi_{\Gamma}=+1$ and $\chi_M=-1$. Upon irradiation, the corresponding Dirac masses can acquire different signs, so that the total Chern number is determined by

\begin{equation}
    C\sim \frac{1}{2}\left[\mathrm{sgn}(m_{\Gamma})-3\,\mathrm{sgn}(m_M)\right],
\end{equation}

up to the overall Berry curvature convention. The different possible mass configurations at the $\Gamma$ and $M$ points, therefore, naturally allow phases with $C=\pm1$ and $C=\pm2$. The precise phase boundaries, however, require the full lattice Floquet Hamiltonian, which we obtain numerically.

Similar to the $p$-wave case, we next investigate the nonlinear Hall responses in bicircularly driven $f$-wave magnets for $r=1$ and $n=3$. The BCD components, $D_x^\Omega$ and $D_y^\Omega$, are shown in Fig.~\ref{fig:response}(a). In contrast to the nearly equal and opposite BCD components obtained for the representative $p$-wave case, the $f$-wave system exhibits an anisotropy between the two components. This originates from the anisotropic momentum-space
distribution of the Berry curvature and band velocities of the effective 
Floquet Hamiltonian. The corresponding BCP responses, $\chi_{xyy}$ and $\chi_{yxx}$, are shown in Fig.~\ref{fig:response}(b). They develop peaks of opposite sign and unequal magnitude. The magnitude as well as the relative strength of both nonlinear responses can be continuously tuned by varying the bicircular light parameters $r$, $R$, and $\alpha$.

Notably, we find that the nonlinear response vanishes for monochromatic driving and requires bicircular driving in an essential manner to be finite. More generally, the response also disappears for frequency ratios satisfying $n+1=0\pmod{3}$. This selection rule is a direct consequence of the underlying $C_3$ rotational symmetry of the triangular-lattice $f$-wave model. Moreover, both the BCD and BCP responses can be effectively tuned by varying the applied light parameters, $r$, $R$, and $\alpha$.

Finally, we examine the spin texture of the $f$-wave magnet for different relative phases of the bicircular drive, $\alpha=0$, $\pi/4$, and $\pi/2$, as shown in Fig.~\ref{fig:response}(c). We find that varying $\alpha$ continuously redistributes the regions of positive and negative out-of-plane spin polarization, $\langle\sigma_z\rangle$, along the Fermi contours and modifies their momentum-space anisotropy. In particular, the locations and relative extent of the spin-polarized segments change appreciably as $\alpha$ is varied, demonstrating that the relative phase $\alpha$ provides direct control over the momentum-resolved spin texture. The spin texture can be further tuned through the amplitude ratio $r$ and the overall driving amplitude $R$.

\section{Summary and outlook}
\label{sec:summary}

We have demonstrated bicircular light-induced topological phases and transitions in odd-parity magnets. We find that bicircular driving gives rise to a rich Chern number phase diagram with a cascade of topological phase transitions enabled by additional control parameters such as the relative amplitude, frequency ratio, and relative phase of the two light components. Beyond the topological transitions, bicircular light enables finite nonlinear Hall responses associated with the Berry curvature and quantum metric contributions. We further identify an odd-even selection rule in the frequency ratio of the bicircular drive, which governs whether these nonlinear responses are symmetry allowed for the $p$-wave case. A variant of this condition is obtained for the $f$-wave magnet, arising from its more intricate lattice symmetry.

Spectroscopic techniques such as angle-resolved photoemission spectroscopy, which have been widely used to characterize unconventional magnets~\cite{krempasky2024altermagnetic,liu2024absence,ding2024large}, could also provide a useful probe of Floquet-induced topological phases~\cite{merboldt2025observation,choi2025observation,wang2026observation}. Furthermore, our predicted light-driven nonlinear responses may also be experimentally accessible~\cite{mciver2020light}. Overall, our results highlight bicircular light as a promising tool for exploring and controlling unconventional magnetic systems via Floquet engineering.

\section*{Data availability statement}

There are no publicly available research data or software supporting this manuscript. Requests for further information or data should be sent to the authors.

\section*{Acknowledgements}

AN thanks A. Jana and M. Ganguli for related collaborations. MAR acknowledges support from a fellowship of the Indian Institute of Science. AN thanks DST CRG grant (CRG/2023/000114) for support. In preparing this manuscript, we used OpenAI's GPT-5.6 model to assist with code development and analysis. All results were verified by the authors. The authors take full responsibility for the content and conclusions of this work.

\onecolumngrid

\setcounter{equation}{0}
\renewcommand{\theequation}{A\arabic{equation}}
\section{Appendix A: Bicircular driving and Floquet effective Hamiltonian}
\label{sec:floquet_formalism}

Here we provide details of deriving the effective Hamiltonian under bicircular light illumination. The counter-rotating $\Omega$-$n\Omega$ bicircular field is introduced through the Peierls substitution $\bk\rightarrow\bk+\mathbf{a}(\tau)$, where $\tau=\Omega t$ and

\begin{equation}
a_x(\tau)=A_1\cos\tau+A_n\cos(n\tau-\alpha),\qquad a_y(\tau)=A_1\sin\tau-A_n\sin(n\tau-\alpha).
\label{eq:bcl_drive}
\end{equation}

We parameterize the two amplitudes by the total amplitude $R$ and the relative amplitude $r$,

\begin{equation}
A_1=\frac{R}{\sqrt{1+r^2}},\qquad A_n=\frac{rR}{\sqrt{1+r^2}},\qquad r=\frac{A_n}{A_1}.
\label{eq:Rr_param}
\end{equation}

One has $A_1^2+A_n^2=R^2$. The driven Hamiltonian is given by

\begin{equation}
\begin{aligned}
H_p(\bk,\tau)
&=\left[-2t\{\cos(k_x+a_x)+\cos(k_y+a_y)\}-\mu\right]\sigma_0
+\lambda\left[\sin(k_y+a_y)\sigma_x-\sin(k_x+a_x)\sigma_y\right] \\
&\quad +J_p\left[c_p\sin(k_x+a_x)+s_p\sin(k_y+a_y)\right]\sigma_z .
\end{aligned}
\label{eq:driven_pwave}
\end{equation}
We expand the time-periodic Hamiltonian as

\begin{equation}
H_p(\bk,\tau)=\sum_{m=-\infty}^{\infty}H_m(\bk)e^{im\tau},\qquad H_m(\bk)=\frac{1}{2\pi}\int_0^{2\pi}H_p(\bk,\tau)e^{-im\tau}\,d\tau.
\label{eq:fourier_blocks}
\end{equation}

For $i=x,y$, we define

\begin{equation}
f_m^{c_i}=\frac{1}{2\pi}\int_0^{2\pi}\cos[a_i(\tau)]e^{-im\tau}\,d\tau,\qquad f_m^{s_i}=\frac{1}{2\pi}\int_0^{2\pi}\sin[a_i(\tau)]e^{-im\tau}\,d\tau.
\label{eq:fc_fs}
\end{equation}

The corresponding lattice form factors are

\begin{equation}
C_{i,m}(k_i)=\cos k_i\,f_m^{c_i}-\sin k_i\,f_m^{s_i},\qquad S_{i,m}(k_i)=\sin k_i\,f_m^{c_i}+\cos k_i\,f_m^{s_i}.
\label{eq:CS_formfactors}
\end{equation}

The Fourier block can then be written as

\begin{equation}
H_m=d_m^0\sigma_0+\mathbf{d}_m\cdot\bm{\sigma},
\label{eq:Hm_dform}
\end{equation}

with

\begin{equation}
d_m^0=-2t(C_{x,m}+C_{y,m})-\mu\delta_{m0},\qquad d_m^x=\lambda S_{y,m},\qquad d_m^y=-\lambda S_{x,m},
\label{eq:Hm_components1}
\end{equation}

and

\begin{equation}
d_m^z=J_p(c_pS_{x,m}+s_pS_{y,m}).
\label{eq:Hm_components2}
\end{equation}

Since $H_p(\bk,\tau)$ is Hermitian,

\begin{equation}
H_{-m}=H_m^{\dagger},\qquad \mathbf{d}_{-m}=\mathbf{d}_m^*.
\label{eq:Hermiticity_blocks}
\end{equation}

To first order in the high-frequency expansion, the effective Hamiltonian reads

\begin{equation}
H_{\mathrm{eff}}(\bk)=H_0(\bk)+\sum_{m=1}^{\infty}\frac{[H_m(\bk),H_{-m}(\bk)]}{m\hbar\Omega}+\mathcal{O}(\Omega^{-2}).
\label{eq:vanVleck}
\end{equation}

The scalar Fourier components do not contribute to the commutator. Using the Pauli algebra

\begin{equation}
[\mathbf{u}\cdot\bm{\sigma},\mathbf{v}\cdot\bm{\sigma}]=2i(\mathbf{u}\times\mathbf{v})\cdot\bm{\sigma},
\label{eq:Pauli_algebra}
\end{equation}

we obtain

\begin{equation}
[H_m,H_{-m}]=2i(\mathbf{d}_m\times\mathbf{d}_{-m})\cdot\bm{\sigma}.
\label{eq:Floquet_commutator}
\end{equation}

It is convenient to introduce the real quantity

\begin{equation}
\kappa_m(\bk)=i\left(S_{x,m}S_{y,-m}-S_{y,m}S_{x,-m}\right).
\label{eq:kappa_def}
\end{equation}

A direct evaluation gives the effective components

\begin{equation}
d_{\mathrm{eff}}^0=-2t(C_{x,0}+C_{y,0})-\mu,
\label{eq:deff0}
\end{equation}

\begin{equation}
d_{\mathrm{eff}}^x=\lambda S_{y,0}-\frac{2\lambda J_ps_p}{\hbar\Omega}\sum_{m=1}^{\infty}\frac{\kappa_m}{m},
\label{eq:deffx}
\end{equation}

\begin{equation}
d_{\mathrm{eff}}^y=-\lambda S_{x,0}+\frac{2\lambda J_pc_p}{\hbar\Omega}\sum_{m=1}^{\infty}\frac{\kappa_m}{m},
\label{eq:deffy}
\end{equation}

and

\begin{equation}
d_{\mathrm{eff}}^z=J_p(c_pS_{x,0}+s_pS_{y,0})+\frac{2\lambda^2}{\hbar\Omega}\sum_{m=1}^{\infty}\frac{\kappa_m}{m}.
\label{eq:deffz}
\end{equation}

Equations~\eqref{eq:fc_fs}-\eqref{eq:deffz} are used for the finite-amplitude all-harmonic lattice calculation. Although the vector potential contains only $\Omega$ and $n\Omega$, the nonlinear lattice functions generate all Fourier harmonics, and the cutoff in $m$ is increased until the observables converge.

\setcounter{equation}{0}
\renewcommand{\theequation}{B\arabic{equation}}
\section{Appendix B: Odd-Even Harmonic Selection Rules and Nonlinear Response}
\label{sec:selection_rules}

In this Appendix we elaborate on the odd-even selection rule for the nonlinear response in $p$-wave magnets. The bicircular vector potential is written as

\begin{equation}
\mathbf{a}(\tau)=
\left(
A_1\cos\tau+A_n\cos(n\tau-\alpha),
A_1\sin\tau-A_n\sin(n\tau-\alpha)
\right),
\end{equation}

where $\tau=\Omega t$. Under a half-period shift, $\tau\rightarrow\tau+\pi$, the fundamental harmonic always changes sign, whereas the $n$th harmonic acquires a factor $(-1)^n$. Hence, for odd $n$, we have

\begin{equation}
\mathbf{a}(\tau+\pi)=-\mathbf{a}(\tau),
\end{equation}

while for even $n$,

\begin{equation}
\mathbf{a}(\tau+\pi)\neq-\mathbf{a}(\tau).
\end{equation}

For odd $n$, this half-period property imposes a selection rule on the Fourier coefficients

\begin{equation}
f^{c_i}_m=\frac{1}{2\pi}\int_0^{2\pi}\cos[a_i(\tau)]e^{-im\tau}d\tau,
\qquad
f^{s_i}_m=\frac{1}{2\pi}\int_0^{2\pi}\sin[a_i(\tau)]e^{-im\tau}d\tau .
\end{equation}

Since $\cos a_i(\tau)$ is invariant under $\tau\rightarrow\tau+\pi$, whereas $\sin a_i(\tau)$ changes sign, one obtains,

\begin{equation}
f^{c_i}_m=0\quad \text{for odd }m,
\qquad
f^{s_i}_m=0\quad \text{for even }m.
\end{equation}

The momentum-dependent Fourier functions,

\begin{equation}
C_{i,m}=\cos k_i\,f^{c_i}_m-\sin k_i\,f^{s_i}_m,
\qquad
S_{i,m}=\sin k_i\,f^{c_i}_m+\cos k_i\,f^{s_i}_m,
\end{equation}

therefore acquire definite momentum parity, which can be expressed as

\begin{equation}
C_{i,m}(-\mathbf{k})=(-1)^m C_{i,m}(\mathbf{k}),
\qquad
S_{i,m}(-\mathbf{k})=(-1)^{m+1}S_{i,m}(\mathbf{k}).
\end{equation}

Consequently, the zeroth Fourier component has an even scalar part and an odd spin-dependent part,

\begin{equation}
d_0^0(-\mathbf{k})=d_0^0(\mathbf{k}),
\qquad
\mathbf{d}_0(-\mathbf{k})=-\mathbf{d}_0(\mathbf{k}).
\end{equation}

The first-order Floquet correction,

\begin{equation}
H_{1}(\mathbf{k})
=
\sum_{m=1}^{\infty}
\frac{[H_m(\mathbf{k}),H_{-m}(\mathbf{k})]}{m\hbar\Omega},
\end{equation}

is even in momentum because both $H_m$ and $H_{-m}$ acquire the same parity factor. Equivalently, for

\begin{equation}
\kappa_m(\mathbf{k})
=
i\left[
S_{x,m}S_{y,-m}-S_{y,m}S_{x,-m}
\right],
\end{equation}

one has

\begin{equation}
\kappa_m(-\mathbf{k})=\kappa_m(\mathbf{k}).
\end{equation}

Thus, for odd $n$, the effective Hamiltonian retains a reciprocal quasienergy spectrum, with the condition

\begin{equation}
E_s(-\mathbf{k})=E_s(\mathbf{k}).
\end{equation}

The band velocity is therefore odd,

\begin{equation}
v_{s,a}(-\mathbf{k})=-v_{s,a}(\mathbf{k}),
\end{equation}

whereas the Berry curvature and quantum metric remain even,

\begin{equation}
\Omega_s(-\mathbf{k})=\Omega_s(\mathbf{k}),
\qquad
g_{ab}(-\mathbf{k})=g_{ab}(\mathbf{k}).
\end{equation}

The Berry curvature dipole,

\begin{equation}
D_a^{\Omega}
=
\sum_s\int_{\mathrm{BZ}}
\frac{d^2k}{(2\pi)^2}
\left(-\frac{\partial f}{\partial E_s}\right)
v_{s,a}(\mathbf{k})\Omega_s(\mathbf{k}),
\end{equation}

therefore has an odd integrand and vanishes for odd $n$,
\begin{equation}
D_a^{\Omega}=0,
\qquad n\ \text{odd}.
\end{equation}

The same argument applies to the metric dipole response. For even $n$, the relation $\mathbf{a}(\tau+\pi)=-\mathbf{a}(\tau)$ is lost. The Fourier coefficients $f^{c_i}_m$ and $f^{s_i}_m$ can then coexist for the same harmonic $m$, so $C_{i,m}$ and $S_{i,m}$ no longer possess definite momentum parity. The effective Hamiltonian may therefore contain mixed even- and odd-in-momentum terms, resulting in

\begin{equation}
E_s(-\mathbf{k})\neq E_s(\mathbf{k}),
\end{equation}

and the cancellation of opposite momenta is no longer enforced.

\twocolumngrid
\bibliography{ref.bib}

\end{document}